\documentclass[aps,prl,twocolumn,showpacs,superscriptaddress,groupedaddress]{revtex4} 

\usepackage{graphicx}  % needed for figures
\usepackage{dcolumn}   % needed for some tables
\usepackage{bm}        % for math
\usepackage{amssymb}   % for math
\usepackage{standalone}
\usepackage{filehook}
\usepackage{gincltex}
\usepackage{collectbox}
\usepackage{filemod-expmin}
\usepackage{cancel}
\usepackage{amsmath}
\usepackage{tikz}

\begin{document}
	
%\centerline{}
%\widetext

%%%%%%\pagebreak
%\newpage
%\clearpage
%\setcounter{page}{1}
\pagenumbering{arabic}

%\title{Azimuthal conformal projection of hyperconical universe: expansion and first CMB tests}
%\title{Link between the $\Lambda$CDM model and hyperconical universes from an azimuthal locally conformal projection}
\title{Geometric interpretation of the dark energy from projected hyperconical universes}
\author{R.~Monjo} %\affiliation{Department of Theoretical Physics I. Complutense University of Madrid, Spain} 
\affiliation{Department of Algebra, Geometry and Topology, Faculty of Mathematics, Complutense University of Madrid, 28040 Madrid, Spain.\\e-mail: rmoncho@ucm.es}
%\affiliation{Climate Research Foundation (FIC), Madrid, Spain}       % D0 authors (remove the first 3 lines
% of this file prior to submission, they
% contain a time stamp for the authorlist)
% (includes institutions and visitors)
% The origin of the dark energy still remains an open issue. 
\date{\today}

\begin{abstract}
This letter explores the derivation of dark energy from a locally conformal projection of hyperconical universes. It focuses on the analysis of theoretical compatibility between the intrinsic view of the standard cosmology and an adequate transformation of hyperconical manifolds. Choosing some parametric family of locally conformal transformations and taking regional (second order) equality between the Hubble parameter of both theories, it is predicted that the dark energy density is $\Omega_{\Lambda} = 0.6937181(2)$. In particular, we used a radially distorted stereographic projection, the distortion parameter of which is theoretically predicted about $\alpha = 0.325 \pm 0.005$ and empirically fitted as $\alpha = 0.36 \pm 0.02$ ($\chi_0^2 = 562$) according to 580 SNe Ia observations.
\end{abstract}


\maketitle

%3750 > 3478 + 205 (150/110*150) = 3683
\section{A. Introduction}
\label{sec:introduction}

\subsection{A.1. Motivation}
\label{sec:motivation}

Astronomical observations suggest that the universe is spatially flat, accelerating and composed of predominately dark energy and dark matter. However, despite of the great agreement of the standard cosmology with this, the nature of dark energy still remains an open issue. In fact, it is not measured from model-independent observations. Moreover, observations of the universe's flatness are also based on the validity assumption of the current model. Changing the frame theory, the same observations can be explained using another geometry, even with positive curvature \cite{Jimenez2009}. 

 %invariance of the geometry under any local conformal transformations
 Recently, geometrical interpretations for the dark energy have been explored \cite{Mannheim2006, Mannheim2007, Maia_et_al_2005}. For instance, conformal gravity can be used to obtain both dark energy and matter \cite{Mannheim2006}. The hypothesis of the conformal cosmology is based on the invariance of the geometry under any local conformal transformation \cite{Mannheim2007}. Alternatively, \citet{Maia_et_al_2005} obtained dark energy as warp in the universe given by the extrinsic curvature within a Friedmann-Robertson-Walker (FRW) universe embedded into a five-dimensional constant curvature bulk. 
 
 The FRW metric used in the standard $\Lambda$CDM model is obtained from a static manifold with constant spatial curvature and an additional scale factor $a(t)$ to explain the expansion. However, other theories based on varying extrinsic geometries can explain the expansion \cite{Melia1, Milne1}. For example, hyperconical universes produce inhomogeneous metrics compatible with the observed expansion and they locally approach to the flat FRW metric. To be consistent with the $\Lambda$CDM model, it can be assumed as a local perturbation theory in inhomogeneous universes expanding regardless of the matter content \cite{Monjo2017}. On the other hand, physical properties of inhomogeneous spaces are interesting because they can be used to trace dark energy and dark matter \cite{Buchert2008}.
 
 As a continuation of the study presented in \cite{Monjo2017}, this letter explores an interpretation of the dark energy according to locally conformal projections of hyperconical universes. For this purpose, the analysis deeps in its theoretical compatibility with the $\Lambda$CDM model for both at low and high redshift regions, particularly including the first acoustic peak in the Cosmic Microwave Backward (CMB) radiation. Therefore, a quick review of the Standard Model is required to compare with the hyperconical model.

\subsection{A.2. The first CMB acoustic peak}
\label{sec:expansion.test}

    %\textit{FRW metric}. A quick review of the Standard Model is required to compare with the proposed model. Recall that FRW metric is obtained by combining a manifold with constant spatial curvature $K$ and a scale factor $a(t)$, which expands with time $t$. One can suppose the spatial section of the universe is a 3-sphere $S^{3}$ of curvature $K = {1}/{R^{2}}$, or a 3-hyperboloid H${}^{3}$ of curvature $K = -{1}/{R^{2}}$, where $R$ is the radius of curvature (i.e., a general curvature is $K = \pm{1}/{R^{2}}$). The Euclidean space (flat universe) can be taken as the limit as $K$ approaches zero. With this, the square of the line element in comoving spherical coordinates ($t$, $r^{\prime}$, $\theta$ ,$\phi$) is:
    % \begin{eqnarray} \label{eq:difFRW}   
    %ds^2 = dt^2 - a(t)^2\left(\frac{d{r'}^2}{1 - K{r'}^2} + {{r'}^{2}}d{\Sigma}^2\right)
    %\end{eqnarray} 
    %where $d{\it{\Sigma}}^2 := d\theta^2 + sin^2\theta d\phi^2$. Scale factor $a(t)$ represents the expansion of the universe, and hence the Hubble parameter is defined as $H := \dot{a}/a$.

	\textit{Hubble parameter}. The Hubble parameter is defined as $H := \dot{a}/a$. For the $\Lambda$CDM model, it is obtained from the second Friedmann equation according to the matter (${\Omega}_{m}$), radiation (${\Omega}_{r}$) and dark energy (${\Omega}_{\Lambda}$) contents. Assuming zero curvature (${\Omega}_K = 0$) , the Hubble parameter $H_{\Lambda}$ of the Standard Model is: \begin{eqnarray} \label{eq:h_lcdm} 
	H_{\Lambda} = {H_o \sqrt{ {\Omega}_{r} (1+{z})^{4} + {\Omega}_{m} (1+{z})^{3} + {\Omega}_{\Lambda}}}
	\end{eqnarray} 
	
     \textit{Comoving distance}. The physical (proper) radial distance or comoving distance, ${\hat{r}{}'}$, is given by the null geodesic with no angular variations in the considered metric. This distance can be written using the redshift $z$ and the Hubble parameter, $H$, with the scale factor expressed as $a/a_{\hat{o}} = 1/(1+z)$ where $a_{\hat{o}} \equiv 1$ is the current value of $a$. For the $\Lambda$CDM universes, one can find: \begin{eqnarray} \label{eq:proper_dist} 
	{\hat{r}{}'}_{\Lambda}(z) = \sin_K \int^{z}_{0} \frac{d{z}}{H_{\Lambda}(z)} 
	\end{eqnarray}where $\sin_K x := \lim_{\epsilon \rightarrow K} \epsilon^{-1/2} \sin (\epsilon^{1/2} x)$, $K = -\Omega_K H_o^2$; i.e., $\sin_0(x) = x$, $\sin_{+1} x = \sin x$ and $\sin_{-1} x = \sinh x$.

	%https://arxiv.org/pdf/1001.4538.pdf 1.726
	%http://power.itp.ac.cn/~huangqg/Publications/JCAP-Distance%20priors%20from%20Planck%202015%20data.pdf
	%https://physics.stackexchange.com/questions/94181/where-is-radiation-density-in-the-planck-2013-results
   %The ‘acoustic peaks’ due to the oscillations in the photon-baryon fluid
	
	\textit{Sound horizon}. The first CMB peak in the multipole moment is given by the sound horizon $r_S$ at decoupling between baryons and photons. It can be obtained according to the integral of the effective sound speed $c_s(z)$ of the coupled baryon-photon plasma \cite{Hu_and_White_1996, Huang_et_al_2015},
	  \begin{eqnarray} %z_{\textit{max}}
	r_S(z) = \int^{t}_{0} c_s(t) a^{-1} dt = \int^{\infty}_{z} \frac{dz}{H(z)} c_s(z) %where $z$ corresponds to $t$ and $z_{\textit{max}} = \infty$ corresponds to $t = 0$. 
	\end{eqnarray}The value of decoupling redshift $z_{dec}$ is $1090.09 \pm 0.42$ according to the Planck mission \cite{A15}. The square value of $c_s(z)$ is: 
	\begin{eqnarray}
	c_s^2 \equiv \frac{\delta P_{\gamma}}{\delta\rho_{\gamma}+\delta\rho_b} = \frac{1}{3(1+R_s(z))}
	 \end{eqnarray} where $\rho_b$ is the baryon density, $\rho_{\gamma}$ is the photon density and $R_s(z)$ is the baryon-to-photon density ratio, which can be expressed in terms of the CMB parameters as: \begin{eqnarray} \label{eq:baryon-to-photon}  
	 R_s(z) \equiv \frac{3}{4}\frac{\rho_b}{\rho_{\gamma}} = \frac{3\omega_b}{4\omega_{\gamma}}\frac{1}{1+z} 
	 \end{eqnarray} where $\omega_b := \Omega_b h^2$ and $\omega_{\gamma} := \Omega_{\gamma} h^2 = \Omega_m h^2/(1+z_{eq})$ are respectively the baryon and radiation contents, expressed with the reduced Hubble constant $h := H /(100^{ }\mathrm{km}^{ }{ }^{ }\mathrm{s}^{-1} \mathrm{Mpc}^{-1})$.
	 
	 Therefore the sound horizon, i.e., the comoving distance traveled by a sound wave, is:
	 \begin{eqnarray} \label{eq:sound_horizon} 
	 r_S(z_{dec}) = \int^{\infty}_{z_{dec}} \frac{d{z}}{H(z) \sqrt{3(1+R_s(z))}}
	 \end{eqnarray}	 
	 Then, the characteristic angle $\theta_A$ of the first CMB peak location and its corresponding multipole $\ell$ are related by:
	 \begin{eqnarray} %\label{eq:first_peak0}
	 %\theta_A = \frac{r_S(z_{dec})}{{\hat{r}{}'}(z_{dec})} \\
	 \label{eq:first_peak}
	 \ell = \frac{\pi}{\theta_A}  = \pi \frac{r'(z_{dec})}{r_s(z_{dec})}
	 \end{eqnarray}Empirically, the first CMB peak is found at $\ell =  220.0 \pm 0.5$ in agreement with the $\Lambda$CDM model ($\ell_{\Lambda} = 220.9$) \citep{A15}.
	 %https://arxiv.org/pdf/1603.03091.pdf
	 
	 Assuming the CMB measurements of the baryon and photon contents, $\omega_b := \Omega_b h^2 = 0.02230(14)$ and $\omega_{\gamma} := \Omega_{\gamma} h^2 = 2.469(26) \cdot 10^{-5}$ \cite{A15}, it is possible to estimate the baryon-to-photon density ratio $R_s(z)$ for any universe model according to Eq. \ref{eq:baryon-to-photon}. With this, the sound horizon $r_S$ and the multipole $\ell$ only depend on the theoretical Hubble parameter $H(z)$. 
	 
	 Therefore, the next sections are focused on the possible compatibility between the standard Hubble parameter and that obtained by the hyperconical universe for low and high redshifts. The local compatibility was found in \cite{Monjo2017}, but for large distances it is required to apply some global projection to the hyperconical universe (Sec. C). %\ref{sec:choice.proj}). 
	 % and on the empirical value of $R_s(z)$ (Eq. \ref{eq:sound_horizon}).
	 
	  \section{B. Hyperconical universes}
	 \label{sec:metric}
	 
	 \textit{Hypercones}. Let $\mathrm{H}_{\beta\pm}^4 \subset \mathbb{R}_{\pm}^{5} := (\mathbb{R}^{5}\smallsetminus \lbrace \textbf{0} \rbrace, \eta_{\pm})$, with metric $\eta_{\pm} := (+1,-1,-1,-1,\pm 1)$, be embedded hypercones with linear expansion and independent of the matter contents, i.e.
	 \begin{eqnarray} % \smallsetminus \lbrace (0,\vec{0},0) \rbrace
	 \mathrm{H}_{\beta\pm}^4 := \{ (t, \vec{r}, u) \in \mathbb{R}_{\pm}^{5} : t^2-\vec{r}^2 \pm u^2 = \beta_{\pm}^2 t^2\}\end{eqnarray}
	 with constant $ \beta_{\pm}^2 \in \{ \beta^2 \in \mathbb{R}: \pm(\beta^2 - 1) > 0 \}$ and $\vec{r} \in \mathbb{R}^{3}$. Alternatively, it can be considered a Wick rotation for the coordinate $u$ with $\eta_{-}$ and $\beta^2 < 1$. In any case, there must exists some diffeomorphism \begin{eqnarray}
	 f:(\mathrm{H}_{\beta\pm}^4, \eta_{\pm})  \rightarrow (\mathbb{R}_{>0} \times \mathbb{R}^3, g) := \mathbb{R}^{1,3}_{g} \end{eqnarray}
	 such that the metric $g$ inherits properties of $\mathrm{H}_{\beta\pm}^4$ and produces the same proper time in $\mathbb{R}^{1,3}_{g}$ as in the Minkowski spacetime.
	 
	 \textit{Deformation}. To preserve the proper time, the initial reference instant $t_{\hat{o}} \in \mathbb{R}_{\geq 0}$ is fixed by the observer although its actual position $(t, \vec{r}, u)$ is moving with time due to the expansion. That is, an expanding deformation $\mathcal{T}_{t}$ is applied to the spatial components $s := (\vec{r}, u) \in \mathbb{R}^{4}$ of the observer according to:
	 \begin{eqnarray} \label{eq:deformation}   
	 \mathcal{T}_{t}: \left(t_{\hat{o}}, s\right) = \left(t_{\hat{o}},  \frac{t}{t_{\hat{o}}} s \right) \in \mathbb{R}_{\pm}^{5} \hspace{1cm} \forall t_{\hat{o}}, t \in \mathbb{R}_{>0}	\end{eqnarray}
	 
	  %For a particular observer the scalar curvature is $R = -6k/t^2$, as is expected for a 3-spherical surface embedded into $\mathrm{R}^{1,4}$. Taking the normalisation $k := 1$, this Ricci scalar coincides for a flat FRW metric with linear expansion $a = t/t_{\hat{o}}$. It has the same effect as if there is no curvature. This effect implies that total density of the universe is equal to the critical density at any time (``flatness problem''). 
	  
	 \textit{Metric}. From the above hypotheses, a radially inhomogeneous metric $g$ is obtained with the same local Ricci curvature as the flat FRW metric \cite{Monjo2017}. The non-zero elements in comoving polar coordinates $(t, r', \theta, \phi)$ are $g_{00} = 2k^{-1}(b-1)+1$, $g_{r'r'} = -a^2/b^2$, $g_{0r'} = -ar'/t_{\hat{o}}b$,  $g_{\theta \theta} = -a^2 {r'}^2$ and $g_{\phi \phi} = -a^2{r'}^2 \sin^2 \theta$; where $k^{-1} := 1-\beta_{\pm}^2$ is the spatial curvature, $a(t) := t/{t_{\hat{o}}}$ is the scale factor and $b(r')^2 := 1 - k{r^{\prime}}^2/t_{\hat{o}}^2$ is an auxiliary function. The diagonal version of the hyperconical metric is given by the coordinate change $t^{\prime} := t\sqrt{2k^{-1}(b-1)+1}$, which is equivalent to selecting $g_{00}^{\prime} = 1$, $g_{0r}^{\prime} = 0$.
	 
	 \textit{Hubble parameter}. For the hyperconical model, the scale factor $a$ is initially defined as linear by hypothesis, and thereby it does not depend on the matter-energy contents. The Hubble parameter is derived using the diagonal coordinates as $H_{hyp} := a^{-1}\partial a/\partial t'$, i.e.
	 \begin{eqnarray} \label{eq:h_hyp} 
	 H_{hyp} = \frac{1}{t'} =  \frac{1}{t_{\hat{o}}}\frac{a_{\hat{o}}}{a} = \frac{1+z}{t_{\hat{o}}}
	 \end{eqnarray}
	 
	  \textit{Comoving distance}. In the hyperconical model, two comoving measures are distinghished, the radial coordinate ${r'}_{hyp}$ of the null geodesic under the extrinsic view and the physical (proper) radial distance ${\hat{r}{}'}_{hyp}$ obtained by projection to the intrinsic view of the hyperconical universe. That is:
	 \begin{eqnarray} \label{eq:proper_dist2}
	 {r'}_{hyp}(z) = \frac{1}{H_o} \xi_k^{^{-1}}( \ln(1 + z) ) \\
	 \label{eq:proper_dist3}
	 {\hat{r}{}'}_{hyp}(z) := f_{\hat{r}}^{\alpha}(t_{\hat{o}},{r'}_{hyp}) 
	 \end{eqnarray}where $f_{\hat{r}}^{\alpha}$ is a projection map (detailed in Sec. C) and the function $\xi_k$, obtained from the null geodesic curve under the diagonal hyperconical metric, is given by \begin{eqnarray} \label{eq:comoving.distance2} 
	 \xi_{k} \left( \frac{r^{\prime}}{t_{\hat{o}}} \right) := \int_{0}^{r^{\prime}} \frac{\sqrt{1-k^{-1}(1-b)^2}}{b(2k^{-1}(b-1)+1)} \frac{d r'}{t_{\hat{o}}}
	 \end{eqnarray} where $b = b(r') := \sqrt{1 - k{r^{\prime}}^2/t_{\hat{o}}^2}$.
	 
	 \section{C. Choice of projection}
	 \label{sec:choice.proj} %\smallsetminus \lbrace (0,\vec{0},0) \rbrace
	 
	 The deformation $\mathcal{T}_{t}: \mathrm{H}_{\beta\pm}^4 \rightarrow \mathbb{R}^{5}_{\pm}$ leads to a differential line that provides the metric $g$, but the output of this transformation is in $\mathbb{R}^{5}_{\pm}$ and still has the unobserved spatial $u$-coordinate. Therefore a projection map $f^{\alpha} : \mathbb{R}^{5}_{\pm} \rightarrow  \mathbb{R}^{1,3}_g$ is required to remove $u$, satisfying $f = f^{\alpha} \circ \mathcal{T}_{t}$, i.e. \begin{center}
	 	%	\textsc{
	 	\begin{tikzpicture}[node distance=1.5cm, auto]
	 	\node (A) {$\mathrm{H}_{\beta\pm}^4$};
	 	\node (B) [right of=A] {$\mathbb{R}^{5}_{\pm}$};
	 	\node (C) [below of=A] {$\mathbb{R}^{1,3}_g$};
	 	\draw[->](A) to node {$\mathcal{T}_{t}$}(B);
	 	\draw[->](A) to node [left] {$f$}(C);
	 	\draw[->](B) to node [right] {$f^{\alpha}$}(C); %[below=0.5ex]
	 	%\node at (1.0099,-1.002) {$\circ$};
	 	\end{tikzpicture}
	 \end{center}	
	 In this work, a family of projection maps $\lbrace f^{\alpha} \rbrace_{\alpha}$ is tested. Each map is applied to comoving coordinates $(t, \vec{r}', u')$ as: \begin{eqnarray}f^{\alpha}: (t, \vec{r}', u) \rightarrow (f^{\alpha}_{\hat{t}}(t,\vec{r}'), f^{\alpha}_{\hat{r}}(t,\vec{r}')) =: (\hat{t}, \vec{\hat{r}}') \in \mathbb{R}^{1,3}_g \end{eqnarray}and it should satisfy several requirements: 
	 
	 (1) \textit{Projection type}. To preserve the local measurement of proper distances, it must be an azimuthal and locally conformal projection; that is, $f^{\alpha}_{\hat{t}}(t,\vec{0}) \equiv t$, $f^{\alpha}_{\hat{r}}(t,\vec{0}) \equiv 0$ and $f^{\alpha}_{\hat{r}}(t,\vec{\epsilon}) \approx \vec{\epsilon}$ for $\left| \vec{\epsilon} \right| \ll t$. Therefore, it is spatially isotropic $f^{\alpha}_{\hat{r}}(t,\vec{r}') = f^{\alpha}_{\hat{r}}(t,r') \vec{r}'/r'$ for all $\vec{r}'$ such as $r':= |\vec{r}'| > 0$.
	 
	 (2) \textit{False boundary}. As $\mathcal{T}_{t}$ is a local projection (done by an observer), it produces an apparent boundary at a certain distance. Particularly, the function $\xi_k(r'/t_{\hat{o}})$ is only real in $r' \in [0, r_k')$ where \begin{eqnarray} \label{eq:rk} 
	 r_k' := \frac{t_{\hat{o}}}{2} \sqrt{4-k}
	 \end{eqnarray} is the boundary of $\mathcal{T}_{t} (\mathrm{H}_{\beta\pm}^4) \ni r'$, where $\xi_k$ diverges. The existence of this limit leads to a divergence in the first derivative of the projection, i.e. $df^{\alpha}_{\hat{r}}(t,r')/dr|_{r'=r_k'} = \infty$.%, since it is a representative of the temporal singularity $t = 0$.
	 
	 (3) \textit{Corrected boundary}. To find the correction of the map $f^{\alpha}$, it must be taken into account that the intrinsic measurement of $|\vec{\hat{r}}'|$ in $\mathbb{R}^{1,3}_g$ is limited by the maximum arc length in $\mathrm{H}_{\beta\pm}^4$, that is $\pi \nu t_{\hat{o}} = \max(\gamma'\nu t_{\hat{o}}) = \max(|\vec{\hat{r}}'|) =  f^{\alpha}_{\hat{r}}(t_{\hat{o}},r_k')$.
	 
	 Unfortunately, the possible projection map $f^{\alpha}_{\hat{r}}$ that satisfies conditions (1) and (2) is not unique. For instance, we tried a distorted stereographic projection %$f^{\alpha}_{\hat{r}} \sim \nu t \mathit{\Gamma}^{\alpha}$ with 
	 \begin{eqnarray}
	 \label{eq:distortion0} 
	  {}_af^{\alpha}_{\hat{r}}(t_{\hat{o}},r_k') \sim \nu t_{\hat{o}}\gamma' {\Delta}^{\alpha}(\gamma'/\gamma_k') \\
	  \label{eq:distortion} 
	  {\Delta}^{\alpha}(x) := \frac{1}{\left(1-x\right)^{\alpha}}
	 \end{eqnarray} where $\gamma' := \gamma'(r') = \sin^{-1}(r'/\nu t_{\hat{o}})$, $\gamma_k' := \gamma'(r_k')$ and ${\Delta}^{\alpha} : [0,1) \rightarrow [0, \infty)$ is a stereographic projector radially distorted by the parameter $\alpha \in \mathbb{R}_{\geq 0}$ and applied to a normalized height $x \in [0,1)$. For an observer, the intrinsic measurement of the distances (including the height) is locally given by the arc length ($\nu t_{\hat{o}}\gamma'$) in $\mathrm{H}_{\beta\pm}^4$.  
	 %\mathit{}
	 
	 Adding the condition (3), Eq. \ref{eq:distortion0} is only valid for a small region ($\gamma' << 1$) because it diverges when $\gamma' \rightarrow \gamma_k'$, and then it requires eliminating this divergence. For example, we tried the inverse stereographic projection \begin{eqnarray} \label{eq:diffeomorphism}  
	 {}_bf^{\alpha}_{\hat{r}}(t_{\hat{o}},r') \sim 2\nu t_{\hat{o}} \tan^{-1} \frac{\gamma'{\Delta}^{\alpha}(\gamma'/\gamma_k')}{2} %\frac{\gamma'/2}{\left(1-\frac{\gamma'}{\gamma_k'}\right)^{\alpha}} 
	%{\hat{r}}' = f^{\alpha}_{\hat{r}}(t_{\hat{o}},r';\alpha) \approx 2\nu t_{\hat{o}} \tan^{-1} \frac{\gamma'/2}{\left(1-\frac{\gamma'}{\gamma_k'}\right)^{\alpha}} 
	% \nonumber \\
	 % \approx 2\nu t_{\hat{o}} \tan^{-1} \frac{\gamma}{\left( 1-\frac{r'}{r_k}\right)^{\alpha} } \vec{k}_{r'} 
	 \end{eqnarray}That is, the projection $f^{\alpha}_{\hat{r}}$ remains one-parametric ($\alpha$). For the temporal projection, it is supposed that $\hat{t} = f^{\alpha}_{\hat{t}}(t,\vec{r}) \approx t$. %1/2-3/4Om = alpha 
	 %and $\gamma= \gamma(r) := \tan^{-1}(r/u(r)) \in [0, \gamma_k)$, and $\alpha$ determines each mapping of the projection family.
	 % That is, the projection is similar than a stereographic type but with $\gamma_k$ instead of $\pi$.
	 
	 %The value of $\alpha$ was fitted using Type Ia supernovae from the Supernova Cosmology Project (SCP) Union2.1 database \citep{B10} \citep{Suzuki2012}. For this purpose, the predicted distance modulus was compared with the observed one. 
	 %In addition, a possible variation of $\alpha$ with $\gamma'$ was analyzed modifying $f^{\alpha}$ with a hypothetical smooth function $\alpha := \alpha(\gamma') \in C^{\infty}$.
	 %\alpha(\gamma(r'(z)))

\section{D. Theoretical compatibility}
\label{sec:compatibility}

\subsection{D.1. Equivalent proper distances}
\label{sec:equivalent.views}

The $\Lambda$CDM and hyperconical models can be respectively interpreted as the intrinsic and extrinsic views of a same local universe, i.e. when distance approaches to zero. This local compatibility between the flat FRW metric and the hiperspherical extrinsic metric (hyperconical universe) leads to an equivalence of the proper or comoving distance obtained according both models when only the intrinsic view is used. Specifically,
\begin{eqnarray} \label{eq:equivalence}
\int^{z}_{0} \frac{d{z}}{H_{\Lambda}(z)} =: {\hat{r}{}'}_{\Lambda}(z) \approx {\hat{r}{}'}_{hyp}(z) =: \int^{z}_{0} \frac{d{z}}{\hat{H}(z)}
\end{eqnarray} for $z << 1$, with the corresponding curvatures ($K = 0$ for $\Lambda$CDM and $k \equiv 1$ for the hyperconical model). In other words, the goal is to find the projection map $f^{\alpha}_{\hat{r}}$ that satisfies a second order (regional) equivalence between an apparent flat Hubble parameter $\hat{H}(z)$ (derived from the projected hyperconical universe) and the measured flat Hubble parameter (according to the $\Lambda$CDM model). To analyze the theoretical regional compatibility between both theories, the function $\hat{H}(z)$ is obtained and compared with $H_{\Lambda}$.
From the right side of the Eq.~\ref{eq:equivalence} and considering Eqs.~\ref{eq:proper_dist2} and \ref{eq:proper_dist3}, it is easy to find that  \begin{eqnarray} \label{eq:local.hubble}
\hat{H}(z) = \left( \frac{d}{dz'} \circ f^{\alpha}_{\hat{r}} \circ \xi_k^{-1} \circ \int_0^{z'} dz \frac{1}{H_{hyp}}  \right)^{-1} (z)
\end{eqnarray} where it is taken as $t_{\hat{o}} \equiv 1$ and thereby $H_{hyp} (z) = 1+z$. The function $\xi_k$ is that provided by Eq.~\ref{eq:comoving.distance2} and the projection map $f^{\alpha}_{\hat{r}}$ is supposed equal to either ${}_af^{\alpha}_{\hat{r}}$ (Eq.~\ref{eq:distortion0}) or ${}_bf^{\alpha}_{\hat{r}}$ (Eq.~\ref{eq:diffeomorphism}). To distinguish both cases, the related Hubble parameters are denoted as ${}_a\hat{H}(z)$ and ${}_b\hat{H}(z)$.

\subsection{D.2. Second order compatibility analysis}
\label{sec:compatibility2}

Because the standard $\Lambda$CDM model has shown good results even for high redshifts, the theoretical compatibility of a hypothetical projection map (Eq.~\ref{eq:diffeomorphism}) should be considered not only at the first order. Expanding the theoretical expression of $H_{\Lambda}$ (Eq.~\ref{eq:h_lcdm}) in terms of Taylor series up to second order, $H_{\Lambda}^{(2)}$, it is found that \begin{eqnarray} \label{eq:hubble_lcdm}  
H_{\Lambda}^{(2)} =  % H_{\Lambda}(0) + \frac{\partial}{\partial s} H_{\Lambda}(s) |_{s=z} z + \frac{1}{2!} \frac{\partial^2}{\partial s^2} H_{\Lambda}(s) |_{s=z} z^2 \nonumber + O(z^3) \\
 \sqrt{\Omega_{r} + \Omega_{m} + \Omega_{\Lambda}} %+ \nonumber \\
+ \frac{4 \Omega_r + 3 \Omega_m}{\sqrt{\Omega_{r} + \Omega_{m} + \Omega_{\Lambda}}} \frac{z}{2} \nonumber \\
+ \frac{8 \Omega_r^2 + (24 \Omega_{\Lambda} + 12 \Omega_{m}) \Omega_{r} + 12 \Omega_{m} \Omega_{\Lambda} + 3 \Omega_m^2}{(\Omega_{r} + \Omega_{m} + \Omega_{\Lambda})^{3/2} } \frac{z^2}{8}
\end{eqnarray} The Hubble parameter derived from the projected hyperconical universe (Eq.~\ref{eq:local.hubble}), expanded up to second order provides: \begin{eqnarray} \label{eq:hubble_hyp}  
%\hat{H}^{(2)} = \frac{1}{\sqrt{k}} + \frac{r_k - 2\alpha}{r_k\sqrt{k}} z + \nonumber \\
%+ \frac{(k+4)r_k^2 + 10\alpha^2 - 6\alpha - 4\alpha r_k^2}{\sqrt{k}\gamma_k^2}\frac{z^2}{4}
  \hat{H}^{(2)} = 1 + \frac{\gamma_k - 2\alpha\sqrt{k}}{\gamma_k} z + \nonumber \\
 + \frac{5\alpha^2k - 2\alpha\sqrt{k}\gamma_k - 3\alpha k + m\gamma_k^2}{2\gamma_k^2} {z^2}
% + \frac{10\alpha^2k - 4\alpha\sqrt{k}\gamma_k - 6\alpha k + 5\gamma_k^2}{\gamma_k^2}\frac{z^2}{4}
\end{eqnarray}where either $m=2$ (for $f^{\alpha}_{\hat{r}} = {}_af^{\alpha}_{\hat{r}}$) or $m=5/2$ (for $f^{\alpha}_{\hat{r}} = {}_bf^{\alpha}_{\hat{r}}$), with $\alpha$ assumed as approximately constant at second order.

Therefore, the equivalence expressed in the Eq.~\ref{eq:equivalence} leads to three equations corresponding to the zero, first and second orders of $H_{\Lambda}^{(2)}$ and $\hat{H}^{(2)}$. The zeroth order corresponds to the trivial flat FRW universe $\Omega_{r} + \Omega_{m} + \Omega_{\Lambda} = 1$. The first order is dependent on the family shape parameter $\alpha$. Even taking the normalization $k \equiv 1$ and assuming that $0 \approx \Omega_{r} << 1$, there are still three unknowns $(\Omega_{m}, \Omega_{\Lambda}, \alpha)$, with two equations. Thus, the second order equality is required to solve the system $H_{\Lambda}^{(2)} = \hat{H}^{(2)}$.

An important point for the reader is that the corresponding system up to third order, $H_{\Lambda}^{(3)} = \hat{H}^{(3)}$, is incompatible with a constant $\alpha$ and $\Omega_{r} \approx 0$. Thereby, the Eq.~\ref{eq:distortion0} and Eq.~\ref{eq:diffeomorphism} should be considered just as second order approximations with constant $\alpha$. 

%{OL = .7182335130, OM = .2817664874}

\subsection{D.3. Explicit second order solution}
\label{sec:explicit.local}

The explicit solution of the second order system $H_{\Lambda}^{(2)} = \hat{H}^{(2)}$, given by Eqs. \ref{eq:hubble_lcdm} and \ref{eq:hubble_hyp}, is:
\begin{eqnarray} \label{eq:explicit_solution}  
\Omega_{\Lambda} = \frac{3\alpha^2 k+2\alpha \sqrt{k} \gamma_k- \alpha k}{2 \gamma_k^2} + \omega_{\Lambda} \\
\Omega_{m} = \frac{2\alpha k(1-3\alpha)}{\gamma_k^2} + \omega_m \\
\Omega_{r} = \frac{9\alpha^2 k-2\alpha \sqrt{k} \gamma_k-3\alpha k}{2\gamma_k^2} + \omega_r
\end{eqnarray} where $\omega_{\Lambda} = 1/2$ or $(6+k)/12$, $\omega_m = 0$ or $-k/3$, $\omega_r = 1/2$ or $(2+k)/4$ depending on whether ${}_a\hat{H}$ or ${}_b\hat{H}$ is used, as well $\Omega_{\Lambda} + \Omega_{m} +\Omega_{r} = 1$. 

%http://hyperphysics.phy-astr.gsu.edu/hbase/Astro/denpar.html
%http://hyperphysics.phy-astr.gsu.edu/hbase/Astro/denpar.html
Considering $k \equiv 1$ and $\Omega_{r} \approx 0$ ($\Omega_{r} = 9.0 \pm 0.5 \cdot 10^{-5}$), the numerical solutions correspond to two complex conjugate sets:
 \begin{eqnarray} \label{eq:solution_valuated}  
\alpha = 0.2830219501(1) \pm c_{\alpha}i\nonumber \\% 0.20427(4)i0.32039(4)i
\Omega_{\Lambda} = 0.6937181(2) \pm c_{\omega}i\\  % 0.26008(3)i 0.40793(3)i
\Omega_{m} = 0.306192(6) \mp c_{\omega}i \nonumber %0.26008(3)i 0.40793(3)i
\end{eqnarray} where $c_{\alpha} = 0.204263(4)$ or $0.320386(2)$, $c_{\omega} = 0.260076(4)$ or $0.407928(3)$ depending on whether ${}_a\hat{H}$ or ${}_b\hat{H}$ is used, and the value in parenthesis is the interval error of last significant figure. Note that the real part of $\Omega_{\Lambda}$ and $\Omega_{m}$ are compatible with the observations updated by the Planck Mission (${\Omega}_{\Lambda} = 0.6911 \pm 0.0062$ and ${\Omega}_{m} = 0.3089 \pm 0.0062$, \citep{A15}). If other positive values of $k$ are considered ($0 < k < 1$), the real values predicted for the parameters remain about $\Omega_{\Lambda} = 0.699 \pm 0.005$, $\Omega_m = 0.3001 \pm 0.005$ and $\alpha = 0.280 \pm 0.003$.

% Figure 1
\begin{figure}
	\includegraphics[width=\columnwidth]{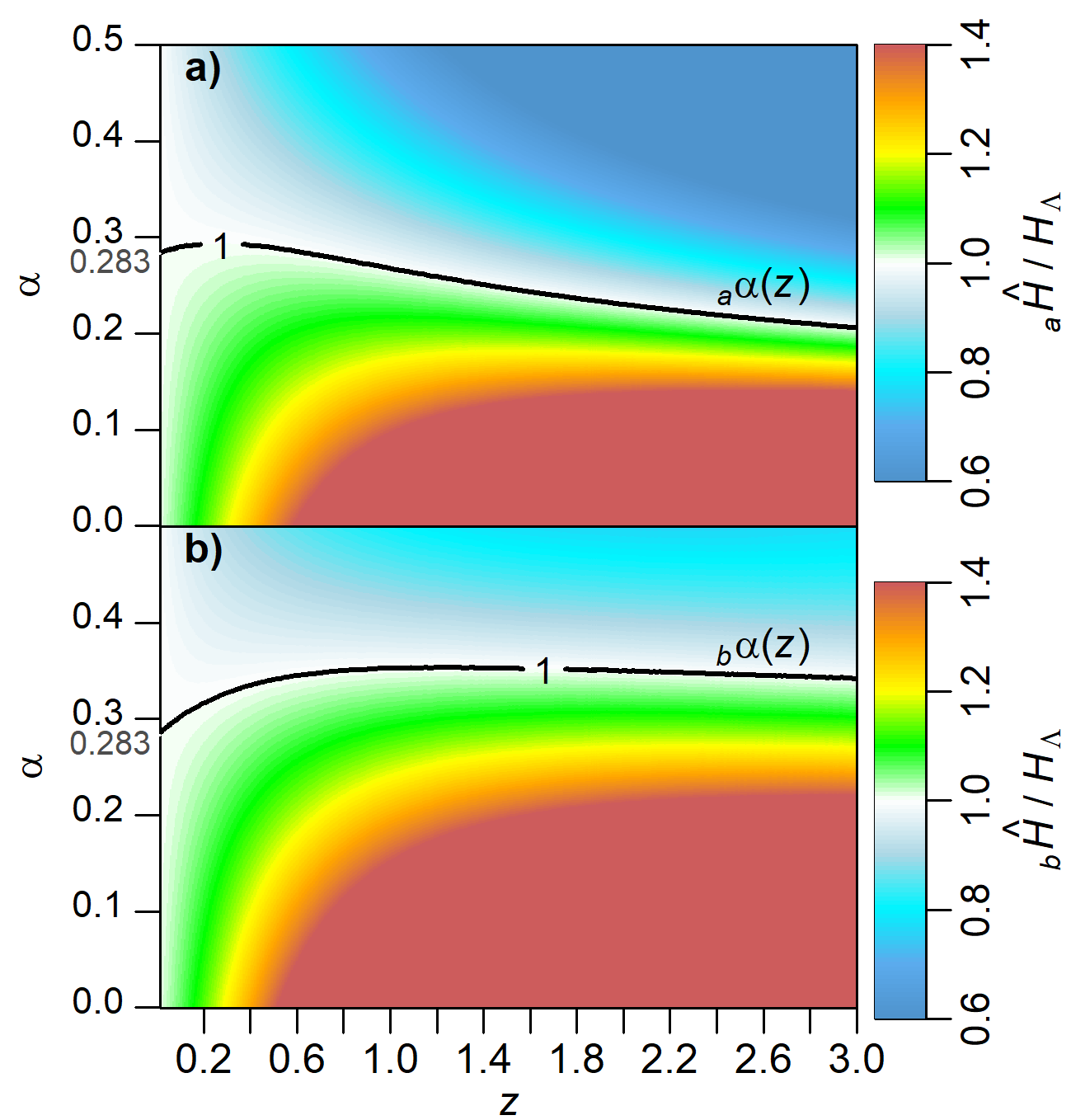} %, natwidth=3779
	\caption{Theoretical compatibility between the Standard Model and the projected hyperconical universe: $a)$ Solution curve ${}_a\alpha(z)$ found for ${}_a\hat{H}=H_{\Lambda}$, $b)$ Solution curve ${}_b\alpha(z)$ found for ${}_b\hat{H}=H_{\Lambda}$%, c) Comparison between the first CMB peak multipole $\ell$ obtained with ${}_bf_{\hat{r}}^{\alpha}$ and the predicted one by the Standard Model, integrating the sound horizon $r_S$ between $z_{max}$ and $z_{dec}$.
	}
	\label{fig:1}
\end{figure}

\textit{\textbf{Proposition}}. Since ${}_af^{\alpha}_{\hat{r}}$ and ${}_bf^{\alpha}_{\hat{r}}$ $\in C^{\infty}[0,r_k')$ lead to the same real part of $\Omega_{\Lambda}$ and $\Omega_m$, there exist smooth \textit{n}-parametric ($\lambda$) transformations $Q_{\lambda}:{}_af^{\alpha}_{\hat{r}} \rightarrow {}_bf^{\alpha}_{\hat{r}}$ such as, for only one configuration of $\lambda$ in each transformation, the imaginary part of these cosmological parameters is zero, i.e. $c_{\omega} = 0$. 

To prove the existence of this unique configuration, it is enough to find the (unique) curves ${}_a\alpha(\gamma')$ and ${}_b\alpha(\gamma')$ that, respectively, solve $c_{\omega} = 0$ for ${}_a\hat{H}(z)$ and ${}_b\hat{H}(z)$. Considering the real part of Eq. \ref{eq:solution_valuated} for $H_{\Lambda}$ and the solution curves ${}_a\alpha(\gamma')$ and ${}_b\alpha(\gamma')$, it is obtained that ${}_a\hat{H}(z) = {}_b\hat{H}(z) = H_{\Lambda}(z)$ (Fig.~\ref{fig:1}), i.e. ${}_af^{{}_a\alpha}_{\hat{r}} = {}_bf^{{}_b\alpha}_{\hat{r}}$. Trivially, there exist parametric maps $q_a(\lambda_a) : {}_af^{\alpha}_{\hat{r}} \rightarrow {}_af^{{}_a\alpha}_{\hat{r}}$ and $q_b^{-1}(\lambda_b) :{}_bf^{{}_b\alpha}_{\hat{r}} \rightarrow {}_bf^{\alpha}_{\hat{r}}$. Therefore, one of the possible transformations is $Q_{\lambda} = Q_{(\lambda_a,\lambda_b)} := q_b^{-1} \circ q_a$ and its unique configuration that solves $c_{\omega} = 0$ is $\lambda = ({\lambda}_a,0)$. $\ensuremath{\blacksquare}$ %, \hfill\ensuremath{\square}  $\hfill\ensuremath{\blacksquare}$
%Figure X shows the curve $\alpha(z) := \alpha(\gamma'(r'(z)))$ for which $\hat{H}/H_{\Lambda} = 1$ with $c_{\omega} = 0$. 

%\textit{\textbf{Consequence}}. The proper distance $\hat{r}_{\Lambda}'$ (Standard Model) can be interpreted as an azimuthal and locally conformal projection of the comoving radial coordinate $r_{hyp}'$ (hyperconical model). 

\subsection{D.4. Agreement with observations}
\label{sec:first.peak}%0.1369 #220 multipole

Empirical values of $\alpha$ were obtained according to 580 Type Ia supernovae (SNe Ia) data, collected from the Supernova Cosmology Project (SCP) Union2.1 database ($z$ from $0.015$ to $1.414$) \cite{Kowalski2008,Suzuki2012}. Theoretical distance modulus was fitted for the projected hyperconical universe (Eq. \ref{eq:proper_dist3}) using ${}_af^{\alpha}_{\hat{r}}$ and ${}_bf^{\alpha}_{\hat{r}}$. For $k \equiv 1$ and considering constant the parameter $\alpha$, the best fit was $\alpha = 0.30 \pm 0.01$ ($\chi_0^2 = 562$) for ${}_af^{\alpha}_{\hat{r}}$ and $\alpha = 0.36 \pm 0.02$ ($\chi_0^2 = 562$) for ${}_bf^{\alpha}_{\hat{r}}$. Within two sigma interval, they are compatible with predicted averages ${}_a\bar{\alpha} = 0.290 \pm 0.005$ and ${}_b\bar{\alpha} = 0.325 \pm 0.005$, estimated for the considered redshift values ($z \leq 1.414$). %Using the regional value of the distortion parameter ()

%, within two sigma interval. Moreover, considering the possible variation of $\alpha$ with $\gamma$, it was found that $|\partial \alpha / \partial \gamma| < 0.005$ and with no significant \textit{p}-values. However, the value of $\alpha = 0.33 \pm 0.01$ ($\chi_0^2 = 563$) fitted for ${}_bf^{\alpha}_{\hat{r}}$ %is not compatible with this theoretical value.  %|\alpha_1| := 

Finally, the projection ${}_bf^{\alpha}_{\hat{r}'}$ can be fitted to obtain the same multipole value for the first CMB peak (Eq. \ref{eq:first_peak}) than the Standard Model (assuming the empirical baryon-to-photon ratio for both models). The distortion parameter that satisfies this equality is $\alpha = 0.320 \pm 0.005$. Therefore, the new model could be compatible with both local and global scales for a distortion of $\alpha = 0.325 \pm 0.005$. %However, the projection proposed in this letter is just an example and another possibility could produce better results.

%H(z) = d/dz f(E(ln(1+z exp(y)-1 )))
%x = exp(z)
%H(exp(y)-1) = d/dz f(E^{-1}(y))
%y := E(r)
%Int( 1/H(z) dz )(exp(E(r))-1)) = f(r)  %easy

\subsection{E. Summary and conclusions}
\label{sec:discussion}%0.1369 #220 multipole

From the hypothesis of a linearly expanding hypersphere (hypercone), it is obtained an inhomogeneous metric that approaches to an effective flat FRW metric at local scale. This is equivalent to a hypothetical compensation
of the matter effect (deceleration) by a dark energy effect (acceleration), at least locally. Consistently, there is a mathematically compatibility between the expansion explained under the Standard Model and under the view of the hyperconical model. 

The (intrinsic) proper distance obtained from the Standard Model could be interpreted as a locally conformal projection of the comoving radial coordinate of the (extrinsic) hyperconical universe. This projection causes a distortion that would provide the origin of the cosmological constant. Particularly, imposing equality in second order, a dark energy density is obtained about $\Omega_{\Lambda} \approx 0.7$. Moreover, it is found a consistent projection (${}_bf^{\alpha}_{\hat{r}'}$), whose predicted distortion parameter for low redshifts ($\alpha = 0.325 \pm 0.005$, $z \leq 1.414$) is compatible with its empirical value ($\alpha = 0.36 \pm 0.02$, fitted to the SNe Ia observations), and very similar to the theoretical value ($\alpha = 0.320 \pm 0.005$) obtained at high redshifts to explain the first CMB acoustic peak. However, the projection proposed in this letter is just an example and other possibilities could produce better results. %This is an example of distortion compatible to explain the nature of the dark energy, but there could be many others.  Particularly, it is predicted that $\Omega_{\Lambda} = 0.6937181(2)$.

%Considering the value of $\Omega_m$ from \ref{eq:solution_valuated},

%Isolating $\alpha$ from Eq. \ref{eq:diffeomorphism}, considering Eq. \ref{eq:proper_dist3} and Eq. \ref{eq:equivalence}, and taking $k \equiv 1$ and $t_{\hat{o}} \equiv 1$, it is obtained: \begin{eqnarray} \label{eq:alpha}  
	% \alpha %= \frac{\ln \left(\gamma(z) \right) - \ln {\hat{r}{}'}_{hyp}(z)  }{\ln \left(1 - \frac{\gamma(z)}{\pi/3} \right)} = \nonumber \\
	% =  \frac{\ln \left(\gamma(z) \right) - \ln \int^{z}_{0} \frac{d{z}}{H_{\Lambda cdm}(z)} }{\ln \left(1 - \frac{\gamma(z)}{\pi/3} \right)}
	% \end{eqnarray} where \begin{eqnarray} \label{eq:gamma}  
	% \gamma(z) = \tan^{-1} \frac{{r'}_{hyp}(z)}{\sqrt{1 - {r'}_{hyp}^2(z)}} \nonumber \\
	%  = \tan^{-1} \frac{\xi_1^{^{-1}}( \ln(1 + z) )}{\sqrt{1 - \xi_1^{^{-1}}( \ln(1 + z) )}}
	% \end{eqnarray} 
	 
\section*{Acknowledgements}

The encouragement and helpful comments received from Prof. R. Campoamor-Stursberg are gratefully acknowledged.
%Thank you for encouraging me to develop these ideas. I would also like to thank Dar\'\i{}o Redolat for encouraging me to continue this work during the last year. Finally, all comments of Prof. Antonio L. Maroto have been very helpful in improving this paper. Otto-Rutwig    % input acknowledgement

\end{document}